\documentclass[twoside,12pt]{article}
\usepackage{epsfig}

\newcommand{\be}{\begin{equation}}
\newcommand{\ee}{\end{equation}}
\newcommand{\bea}{\begin{eqnarray}}
\newcommand{\eea}{\end{eqnarray}}

\begin{document}

\title{ \vspace{1cm} Is Cen A  surrounded by tens EeV Multiplet?}
\author{D.\ Fargion,$^{1,2}$ \\
$^1$ Physics Department, Rome University 1, Ple. A. Moro 2, 00185\\
$^2$INFN, Roma 1, Italy\\}
\maketitle

\begin{abstract}
Ultra High Energy Cosmic Rays (UHECR) at GZK cut off energy ($E\geq 5.5 \cdot 10^{19}$ eV) may keep sharp or diffused directionality wherever their composition is made by nucleon or light nuclei. Auger experiment UHECR (2007-2010) did show a mild clustering around Cen A. Two over three of the recent discovered AUGER multiplet (a dozen of events each) tail clustering at twenty EeV are pointing to primary sources very  near the same UHECR crowded Cen A region. These tens EeV tail is aligned with the same UHECR events. We foresaw such possibility as fragment tails of lightest UHECR nuclei. We discuss the relevance of this correlation within a model where UHECR are mostly lightest He like nuclei.
UHECR fragment multiplet clustering aligned along higher Cen A events (above  $5.5 \cdot 10^{19} $ eV  energy)  probe and reinforce our interpretation with an a priori probability to occur below $3 \cdot 10^{-5}$.
\end{abstract}
\section{Fragments  in flight along UHECR}
 UHECR Astronomy and nuclear composition are more and more in severe conflict. This contradiction were already inherited in  early, apparent, discover of a Super Galactic UHECR correlation \cite{Auger-Nov07}. In that
 key article the needed UHECR directionality (assumed nucleon) have been , at the same time, in disagreement with the first observed nuclei composition.
Indeed UHECR particle astronomy may rise, suffering however of some directionality spread by the smearing of magnetic field bending along the UHECR flight. This bending is negligible for proton and He, but sever for iron.
 The bending maybe coherent (just one direction) if the magnetic field is constant or it is random , if the field directions are changing , as it happens inside the galactic plane along and across the galactic arms while pointing to Cen A.   The nucleon, the light nuclei   may keep more or less precise directionality,   a smearing astronomy, while  heaviest ones may exhibit only tiny  anisotropy if  sources are near.
      In addition to these signals, UHECR in flight are making  fragment secondaries nuclei as well as,  because photo-dissociation, parasite  gamma and neutrino tails somehow correlated.
   These UHECR compositions are leading to different secondary gamma and neutrino spectra; these different nature are making UHECR nucleon origination local and well directed (GZK cut off, tens Mpc distances) or even much local and smeared (a few Mpc) for our lightest UHECR nuclei considered in recent articles \cite{Fargion2008}. Heavy nuclei may also have, by photon-disruption a gamma and neutrino secondary tail, probably so smeared to be mostly sink into background noise: if UHECR are only iron, as some authors still believe,  than UHECR astronomy, in particular the extragalactic one, will be  so much bent, polluted and  smeared to be  hopeless: only large scale iron anisotropy might occur by nearest galactic sources.  On the contrary extragalactic lightest nuclei UHECR astronomy may be  surrounded by a parasite (a little smeared) trace made by point source gamma, neutrinos (TeVs-PeVs) and also tens EeV energy UHECR fragments. We did suggested this possibility since few years \cite{Fargion2011}. The lightest UHECR nuclei model \cite{Fargion2008} is able to explain the surprising absence or paucity of events  toward Virgo (the nearest extragalactic cluster of galaxy) and the angle range $10^{o}-15^{o}$, the spread directionality (orthogonal to galactic arms) of  events around and along Cen A.
   A large number of authors discussed mainly the spectra of UHECR cut off as a key discriminator for UHECR modeling. However the mass composition role may confuse the real shape of any (apparent) GZK cut off.
   We address more on the UHECR anisotropy nature able we hope  to  correlate UHECR maps and composition, using all radio, IR, gamma MeV-GeV-TeV, maps available.

\begin{figure}[htb]
\begin{center}
\epsfig{file=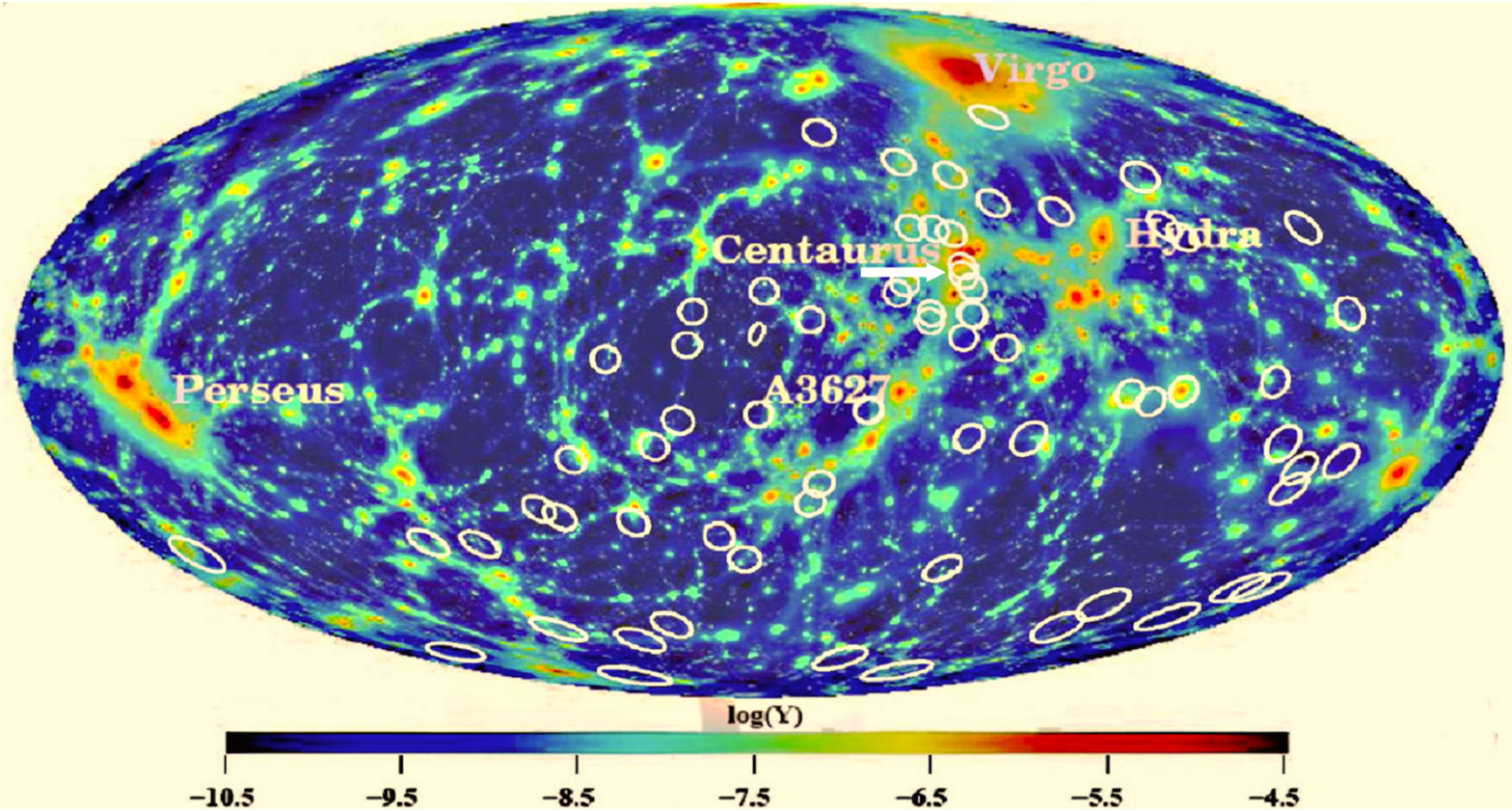,scale=0.16}
\epsfig{file=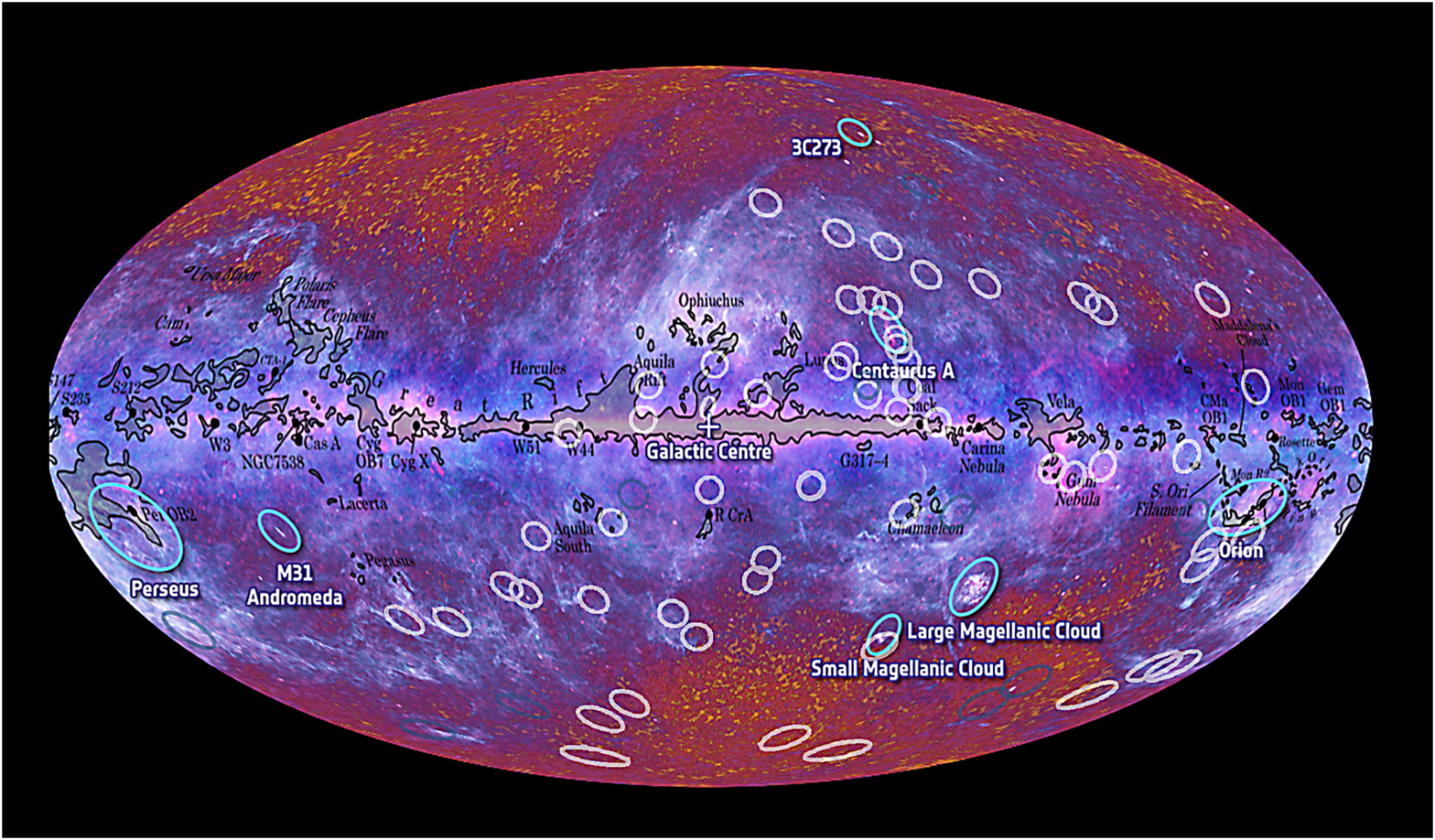,scale=0.15}
\caption{Left: The last 2010 UHECR event map  event map overlap with nearby infrared sky map; the clustering toward Cen A is the main UHECR signal.  Around this source the presence of a twin collinear multiplet clustering at twenty EeV shown in next figures. Note the relevance of Infrared Virgo cluster and its absence in AUGER map. Note also a tiny correlation of Vela with an unique galactic triplet event. Right: The last Planck infrared map and labels , whose spread white noise is due to galactic dust, with UHECR events. Note the partial suppression of UHECR events in those  regions where white dust is missing: this absence may hint for a galactic component of UHECR.}\label{1r}
\end{center}
\end{figure}

\subsection{Deuterium, proton, gamma and neutrino tails }

 \begin{figure}[!t]
  \vspace{5mm}
  \centering
  \includegraphics[width=3.1 in]{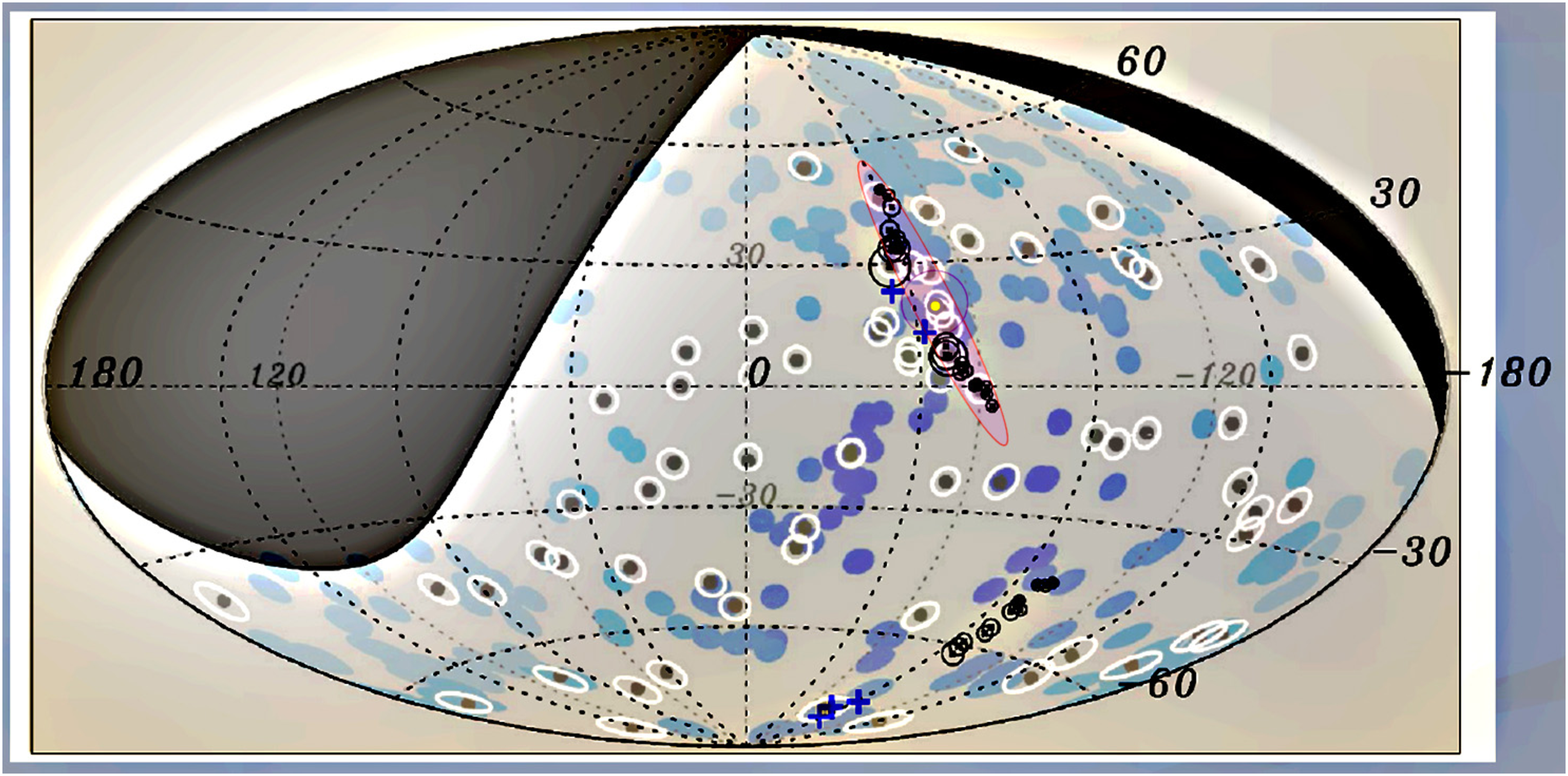}
  \includegraphics[width=2.8 in]{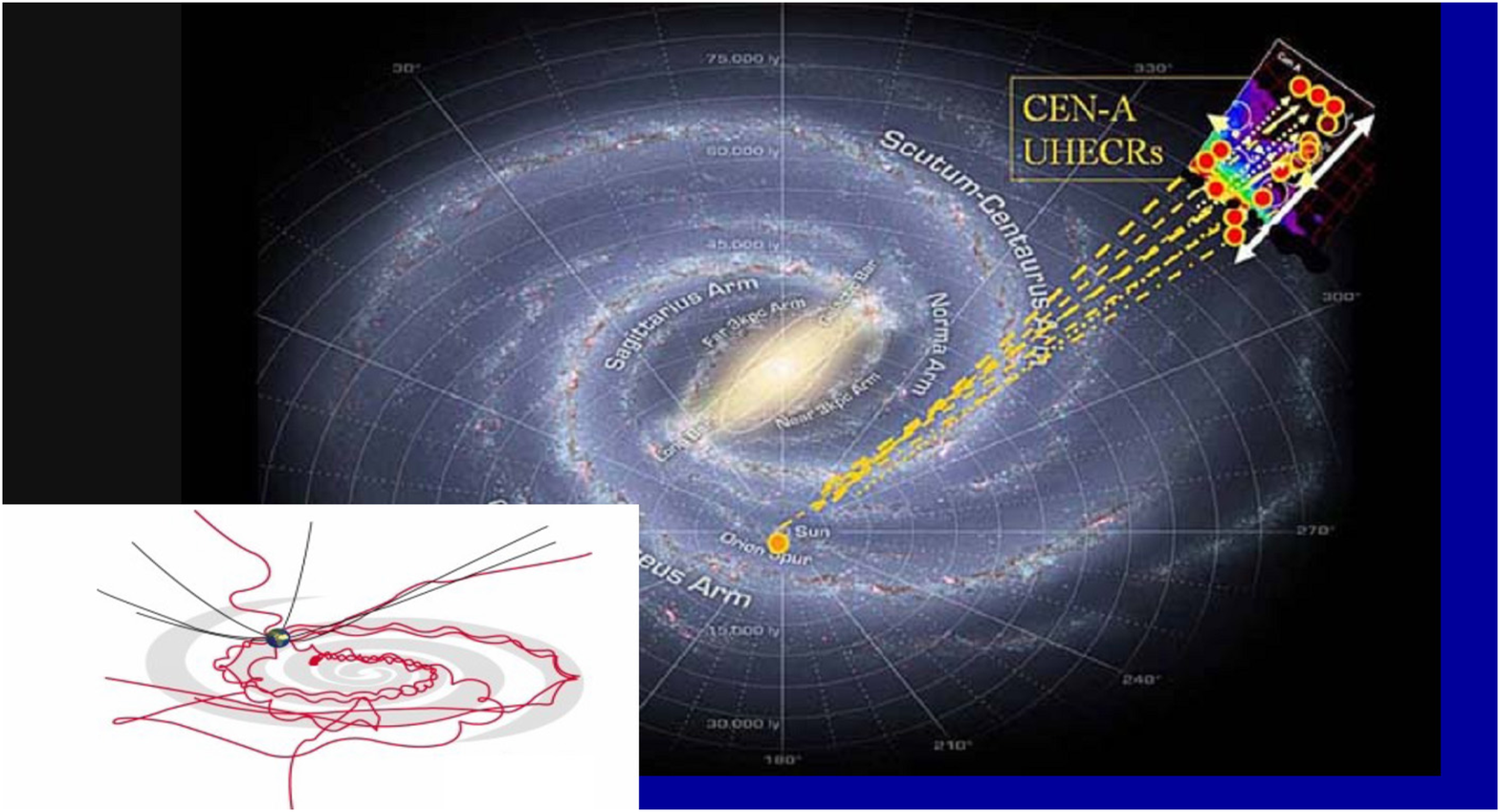}
  \caption{Left: the AUGER 2010 UHECR event map and two of the three multiplet clustering toward Cen A; their sources as shown by dotted curve are  within a tiny disk area (at radius of $10^{o}$); the dotted ellipsoid area of the UHECR and multiplet clustering is also extremely correlated, aligned and small. Right: multiplet clustering toward Cen A by random bending of the galactic  spiral magnetic fields.}
  \label{fig2}
 \end{figure}

 \begin{figure}[!t]
  \vspace{5mm}
  \centering
  \includegraphics[width=2.8 in]{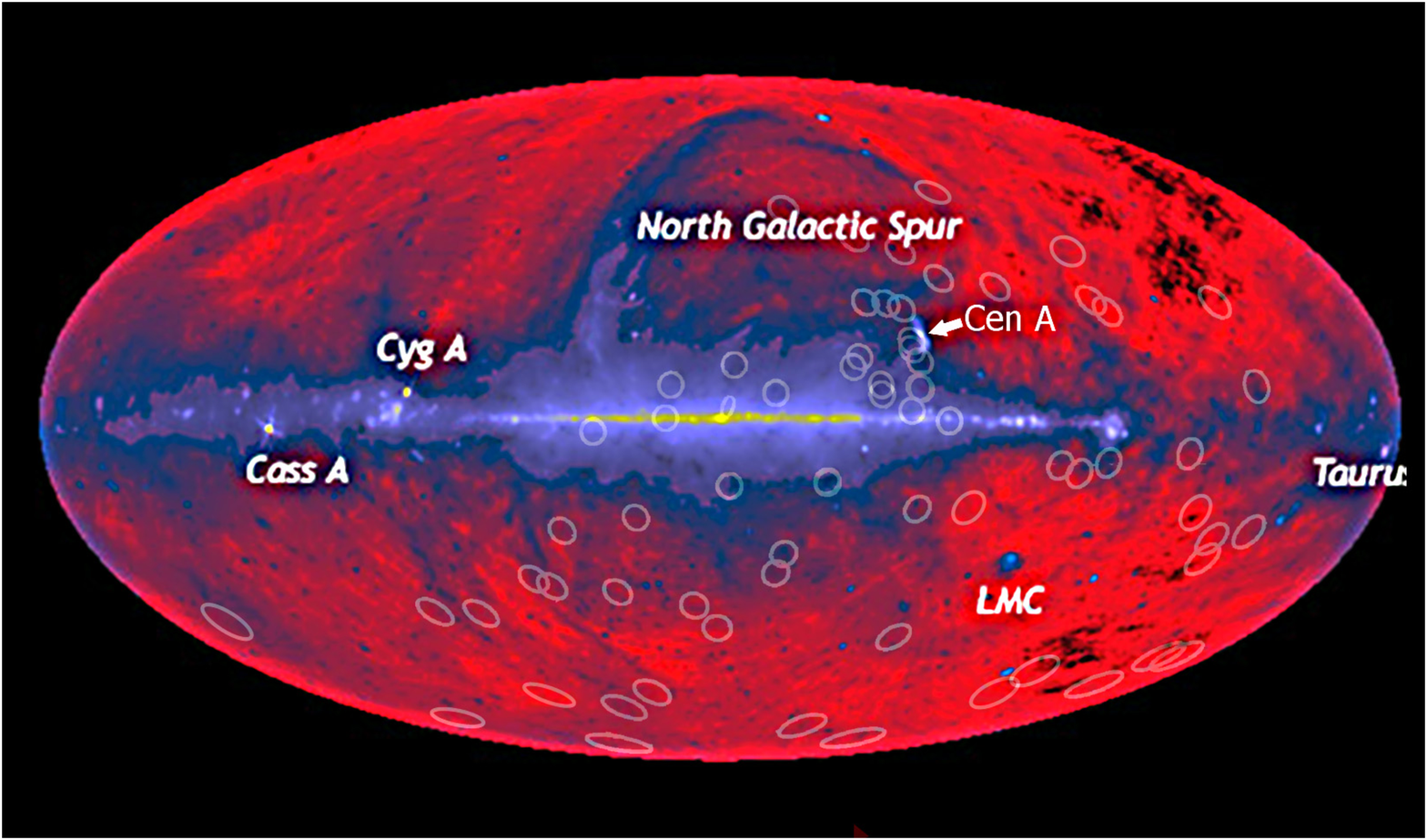}
    \includegraphics[width=3.2 in]{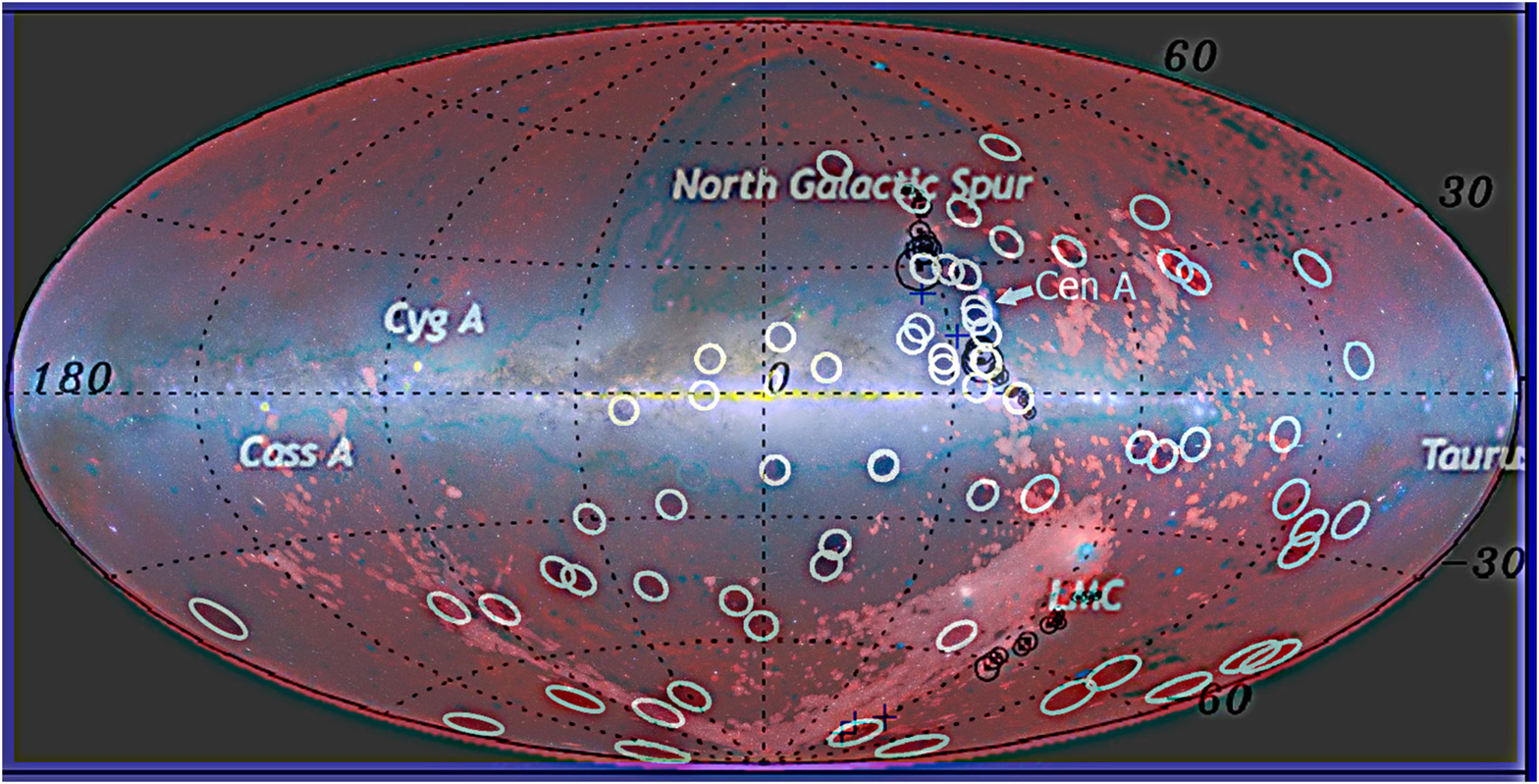}
  \caption{Left: The AUGER 2010 UHECR event map on radio 408 Mhz map and two of the three multiplet clustering toward Cen A. Right:  the third Multiplet clustering points somehow toward Large and Small Magellanic clouds  as well it overlaps with the clustering along the  Magellanic stream. }
  \label{fig2-3}
 \end{figure}

   UHECR formed (mostly) by  lightest nuclei may explain a partial clustering of events, as the one around CenA as well as a puzzling UHECR absence  around Virgo. Light nuclei are fragile and fly few Mpc before being
    halted by photo-disruption \cite{Fargion2008},\cite{Fargion2009},\cite{Fargion2010}. Their fragments $He + \gamma \rightarrow D+D,D+\gamma \rightarrow p+n+ \gamma  , He + \gamma \rightarrow He^{3}+n, He + \gamma \rightarrow T +p $  may nevertheless trace the same UHECR maps by a secondary  clustering at half or even fourth of the UHECR primary  energy \cite{Fargion2011}.  Neutrinos and gamma are tracing  (both for nucleon or light nuclei) their UHECR trajectory, respectively growing at EeVs (if nucleon) or PeV-TeV (if light nuclei) energy. Gamma secondaries rays from cosmic sources are partially absorbed by microwave and infrared background making once again a very local limited UHECR-gamma astronomy. Among neutrinos $\nu$,  muons ones $\nu_{\mu}$, the most penetrating and easy to detect on Earth, are unfortunately deeply polluted by a rich atmospheric  component (as smeared and as the isotropic  parent CR nucleons and nuclei ). This atmospheric  isotropy and homogeneity is probed by last TeV muon neutrino maps in a very smooth ICECUBE neutrino map. Tau neutrinos on the contrary, the last neutral lepton discovered, are almost absent in atmospheric secondaries (about five order of magnitude suppressed at TeVs). Rare $\nu_{\mu}\rightarrow \nu_{\tau} $ neutrino oscillation at ten GeV  atmospheric windows,  may nevertheless arise; at tens TeV-PeV up to EeV  $\nu_{\tau} $ neutrino might be a clean signal of  UHECR-neutrino associated astronomy\cite{FarTau}. Their tau birth in ice  may shine as a double bangs (disentangled above PeV)  \cite{Learned}. In addition UHE tau, born tangent to the  Earth or mountain, while escaping  in air  may lead, by decay in flight, to  loud, amplified  and well detectable tau-airshower at horizons \cite{Fargion1999},\cite{FarTau}. Both in atmosphere fluorescence tracks or by Cherenkov blazing, or by partial skimming ground detectors. Tau astronomy versus UHECR are going to reveal most violent  sky as the most deepest probe. This tau airshowers or Earth skimming neutrino \cite{Feng02} were considered since more than a decade and are going to be observed in AUGER or TA in a few years\cite{FarTau},\cite{Auger-01},\cite{Feng02},\cite{Auger07}.  Regarding the puzzle of UHECR let us also remind that more than a decade ago we were facing and solving a  how obsolete problem due to  AGASA and Fly Eyes events; such events were calling for sources at distances above GZK cut off. The solution was  based on the sources ejecting primary UHE neutrinos at ZeV energy scattering on relic (cosmic) ones making Z  bosons in flight and, after decay,  UHECR nucleons on Earth \cite{Fargion1997}.

\section{Multiplet: Cen A, Magellanic stream and  Galactic sources}

     In recent maps of UHECR we noted first hint of  Vela (see Fig. \ref{fig2-3}), the brightest and nearest gamma source, a first galactic source  is rising as a UHECR triplet nearby \cite{Fargion09b}.
     Since earliest maps we found that Cen A (the most active and nearby extragalactic AGN) is apparently shining  UHECR source whose clustering (almost a quarter of the event) along a narrow solid angle around (whose opening angular size is  $\simeq 17^{o}$) seem  firm and it is  favoring as we mentioned, lightest nuclei \cite{Fargion2008},\cite{Fargion09a}, \cite{Fargion09b}, \cite{Fargion2009}.
     Remaining UHECR events are possibly   heavier nuclei more bent and smeared of  galactic and-or partially  nearby extragalactic origin. As a possible Magellanic stream (see Fig. \ref{fig2-3}) following the third multiplet disposal.
As we mentioned in the abstract UHECR map, initially (2007) consistent with  GZK volumes \cite{Auger-Nov07}, to day (2010 map) seem to be not much correlated with expected Super Galactic Plane \cite{Auger10}. Moreover slant depth data of UHECR from AUGER airshower shape do not favor the proton but points to  a nuclei.  To make even more confusion HIRES, on the contrary, seem to favor, with less statistical weight, UHECR mostly nucleons. We tried  (at least partially) to solve the contradictions assuming UHECR mostly as light nuclei ( $He^4$ and maybe Li, Be) spread by planar spiral galactic fields, randomly , bending them, as observed,  vertically to galactic axis.

\subsection{The Lorentz bending and the  UHECR  and Multiplet}
Cosmic Rays directions are blurred by magnetic fields. Also UHECR suffer of a Lorentz force deviation. More if nuclei, little if nucleon.  As mentioned this smearing maybe source of UHECR features. We see them mostly along Cen A.
 There are  at least three main bending of UHECR along galactic plane. The  extra-galactic events in intergalactic spaces do not suffer much bending. However UHECR in late (local group, galactic) bending feel coherent galactic arm field, as well as random fields due to the  turbulence or random fields along the whole galactic plane, or inside each arms. Let us renumber them\\
(1)
The coherent Lorentz angle bending $\delta_{Coh} $ of a proton (or nuclei) UHECR (above GZK \cite{Greisen:1966jv}) within a galactic magnetic field  in a final nearby coherent length  of $l_c = 1\cdot kpc$ is $ \delta_{Coh-p} .\simeq{2.3^\circ}\cdot \frac{Z}{Z_{H}} \cdot (\frac{6\cdot10^{19}eV}{E_{CR}})(\frac{B}{3\cdot \mu G}){\frac{l_c}{kpc}}$.\\
(2) The random bending by random turbulent magnetic fields, whose coherent sizes (tens parsecs) are short and whose final deflection angle is  smaller than others and they are here ignored.\\
(3) The random, but alternate, multiple UHECR bending along the galactic plane across  alternate arm magnetic field, whose directions are inverted each other (See Fig.\ref{fig2}). Its effect will be ruling for UHECR and multiplet Cen A bending. \\

The corresponding coherent  bending of an Helium UHECR at same energy, within a galactic magnetic field
  in a wider nearby coherent length  of $l_c = 2\cdot  kpc$ is $\delta_{Coh-He} \simeq
{9.2^\circ}\cdot \frac{Z}{Z_{He}} \cdot (\frac{6\cdot10^{19}eV}{E_{CR}})(\frac{B}{3\cdot \mu G}){\frac{l_c}{2 kpc}}
$.

This bending angle  is somehow compatible with observed multiplet along $Cen_A$ (and also the possible clustering along Vela), at much nearer distances; indeed in Vela case it is possible  a larger magnetic field along its direction (20 $\mu G$) and-or  a rare heavy iron composition $\delta_{Coh-Fe-Vela} \simeq {17.4^\circ}\cdot \frac{Z}{Z_{Fe}} \cdot (\frac{6\cdot10^{19}eV}{E_{CR}})(\frac{B}{3\cdot \mu G}){\frac{l_c}{290 pc}}$. Such iron UHECR are mostly bounded inside a Galaxy, as well as in Virgo, explaining partially its extragalactic absence. In lightest nuclei model most of heavier iron nuclei  may be bounded inside  Virgo. We shall not discuss   rare (and in our view improbable) deflection from Virgo tuned and overlayed into Cen A direction as some author suggested  \cite{Semikoz10}.

 The incoherent random angle bending (2) along the galactic plane and arms, $\delta_{rm} $, while crossing along the whole Galactic disk $ L\simeq{20 kpc}$  in different spiral arms  and within a characteristic  coherent length  $ l_c \simeq{2 kpc}$ for He nuclei is $$\delta_{rm-He} \simeq{16^\circ}\cdot \frac{Z}{Z_{He^2}} \cdot (\frac{6\cdot10^{19}eV}{E_{CR}})(\frac{B}{3\cdot \mu G})\sqrt{\frac{L}{20 kpc}} \sqrt{\frac{l_c}{2 kpc}}$$ The heavier  (but still lightest nuclei) bounded from Virgo are Li and Be:
$\delta_{rm-Li} \simeq {24^\circ}\cdot \frac{Z}{Z_{Li^3}} \cdot (\frac{6\cdot10^{19}eV}{E_{CR}})(\frac{B}{3\cdot \mu G})\sqrt{\frac{L}{20 kpc}}
\sqrt{\frac{l_c}{2 kpc}} $, $\delta_{rm-Be} \simeq{32^\circ}\cdot \frac{Z}{Z_{Be^4}} \cdot (\frac{6\cdot10^{19}eV}{E_{CR}})(\frac{B}{3\cdot \mu G})\sqrt{\frac{L}{20 kpc}}
\sqrt{\frac{l_c}{2 kpc}}$.  It should be noted that the present anisotropy above GZK \cite{Greisen:1966jv} energy $5.5 \cdot 10^{19} eV$ might leave a tail of signals: indeed the photo disruption of He into deuterium, Tritium, $He^3$ and protons (and unstable neutrons), might rise as clustered events at half or a fourth (for the last most stable proton fragment) of the energy:\emph{ protons being with a fourth an energy but half a charge He parent may form a tail  smeared around Cen-A at twice larger angle} \cite{Fargion2011}. We suggested  to look for correlated tails of events, possibly in  strings at low $\simeq 1.5-3 \cdot 10^{19} eV$ along the $Cen_A$ train of events. \emph{It should be noticed that Deuterium fragments are half energy and mass of Helium: Therefore D and He spot are bent at same way and overlap into UHECR circle clusters}\cite{Fargion2011}.  Deuterium are even more bounded in a local Universe because their fragility. In conclusion He like UHECR  maybe bent by a characteristic as large as  $\delta_{rm-He}  \simeq 16^\circ$; its expected lower energy Deuterium tails at half energy ($30-25 EeV$)  also at ($\delta_{rm-p}  \simeq 16^\circ$); protons last traces at a quarter of the UHECR energy, around twenty EeV multiplet, will be spread within ($\delta_{rm-p}  \simeq 32^\circ$). \emph{All well within the observed CenA UHECR  clustering and its parasite correlated twin multiplet spread shown in figures.} See  Fig. \ref{fig2},\ref{fig2-3}.

\begin{figure}[!t]
  \vspace{5mm}
  \centering
  \includegraphics[width=2.8 in]{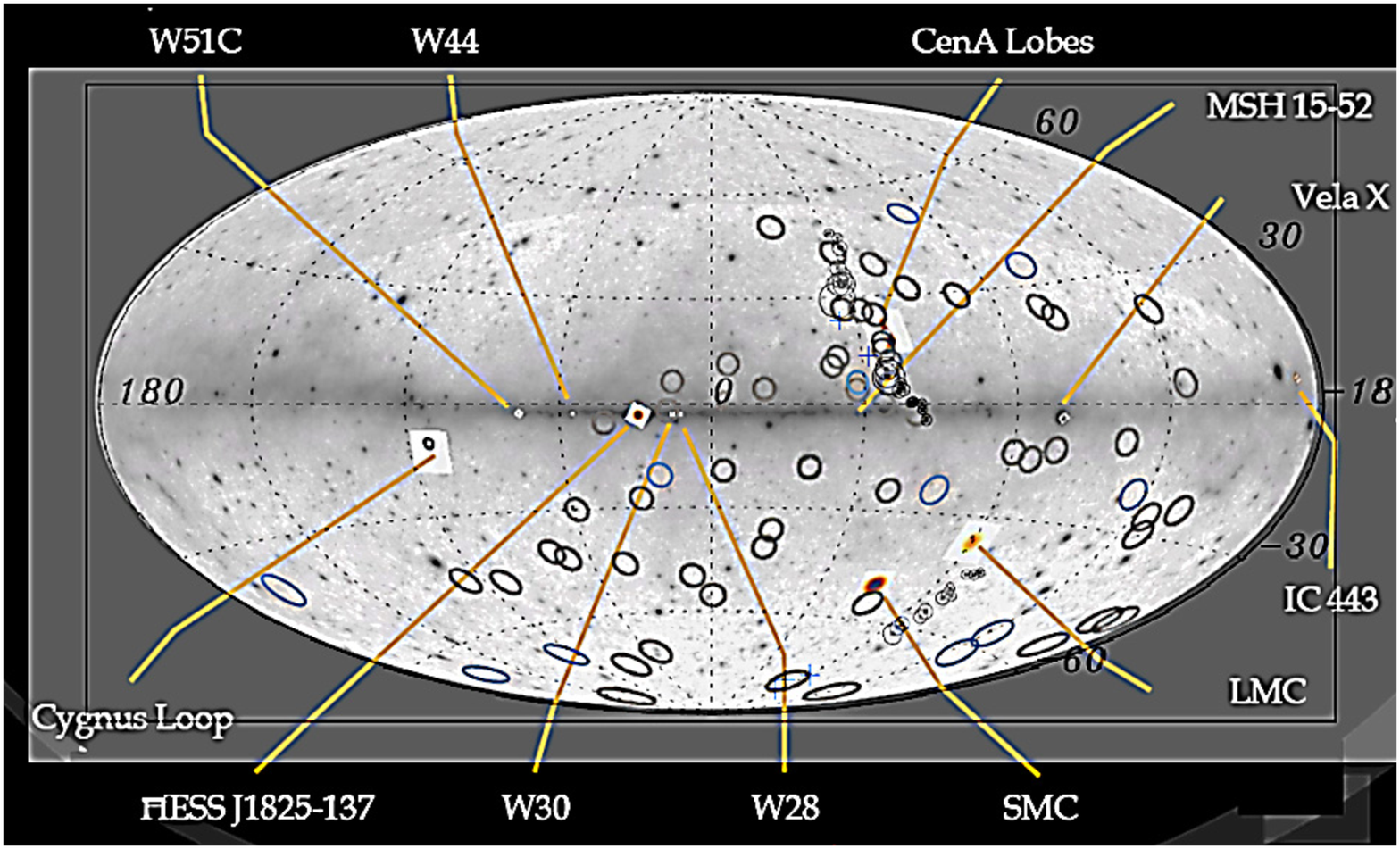}
  \includegraphics[width=3.2 in]{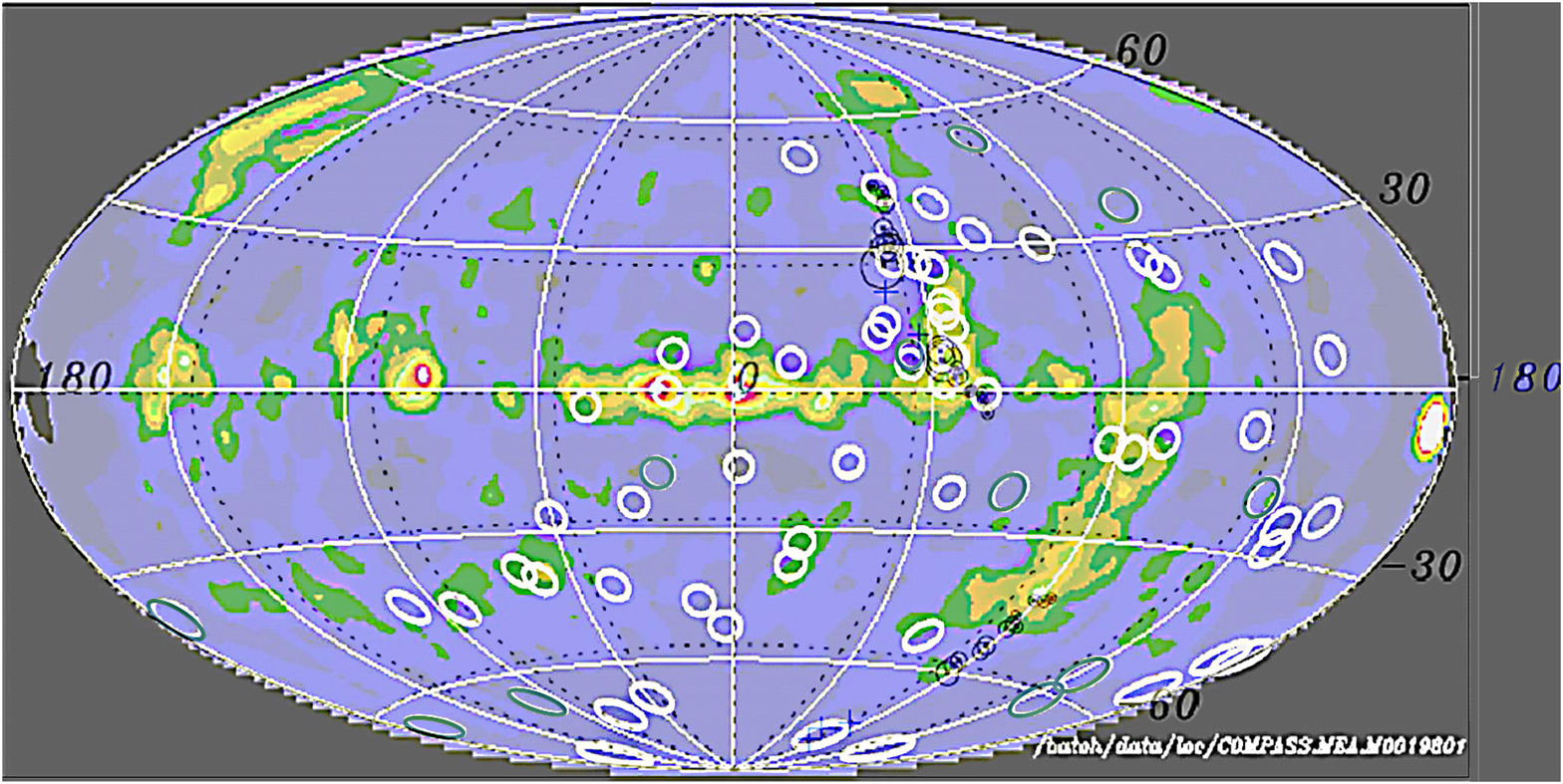}
  \caption{Left: last 2010 UHECR event map by AUGER and the overlap multiplet clustering toward Cen A, inside the last Fermi gamma map and labels, whose twin multiplet expected sources are within a tiny disk area (of  radius below $10^{o}$). Right: last 2010 UHECR event map by AUGER and the Multiplet clustering toward Cen A overlap the MeV Comptel gamma map; note the apparent clustering of UHECR along the Vela, Magellanic stream , Cen A and other galactic regions. These gamma area may contains additional clustering in future records probing a  galactic nature of a fraction of UHECR.}
  \label{fig4-5}
 \end{figure}

 The very  recent multiplet clustering published just few weeks ago by AUGER at twenty EeV  contains just three and apparently isolated train of events  with  (for the AUGER collaboration) no statistical meaning. \cite{Auger11}. Indeed   following that paper   both multiplet head cross are pointing toward two unknown sources, each nearby Cen A (See Fig. \ref{fig4-5},\ref{fig2}) \emph{both bent in opposite way}. This assumption (a magnetic field coherently bending each source in opposite way) just on the boundary along Cen A, seem really unrealistic. To miss the Cen A correlation seem a prejudice.
  
    Indeed the two multiplet sources, as suggested by  \cite{Auger11}, are pointing one down toward the galactic plane and the other up, opposite to the galactic plane. Each train of events is calling for an opposite fine tuned coherent magnetic field. On the contrary our random bending interpretation by unique source originated in Cen A,  occur (naturally) up and down because the well known observed, alternate and horizontal,  zigzag spiral magnetic fields  while UHECR are crossing trough each galactic arm,  flying and skimming  the galactic plane (See inner right figure Fig. \ref{fig2}).
  
\begin{figure}[!t]
  \vspace{5mm}
  \centering
  \includegraphics[width=3.0 in]{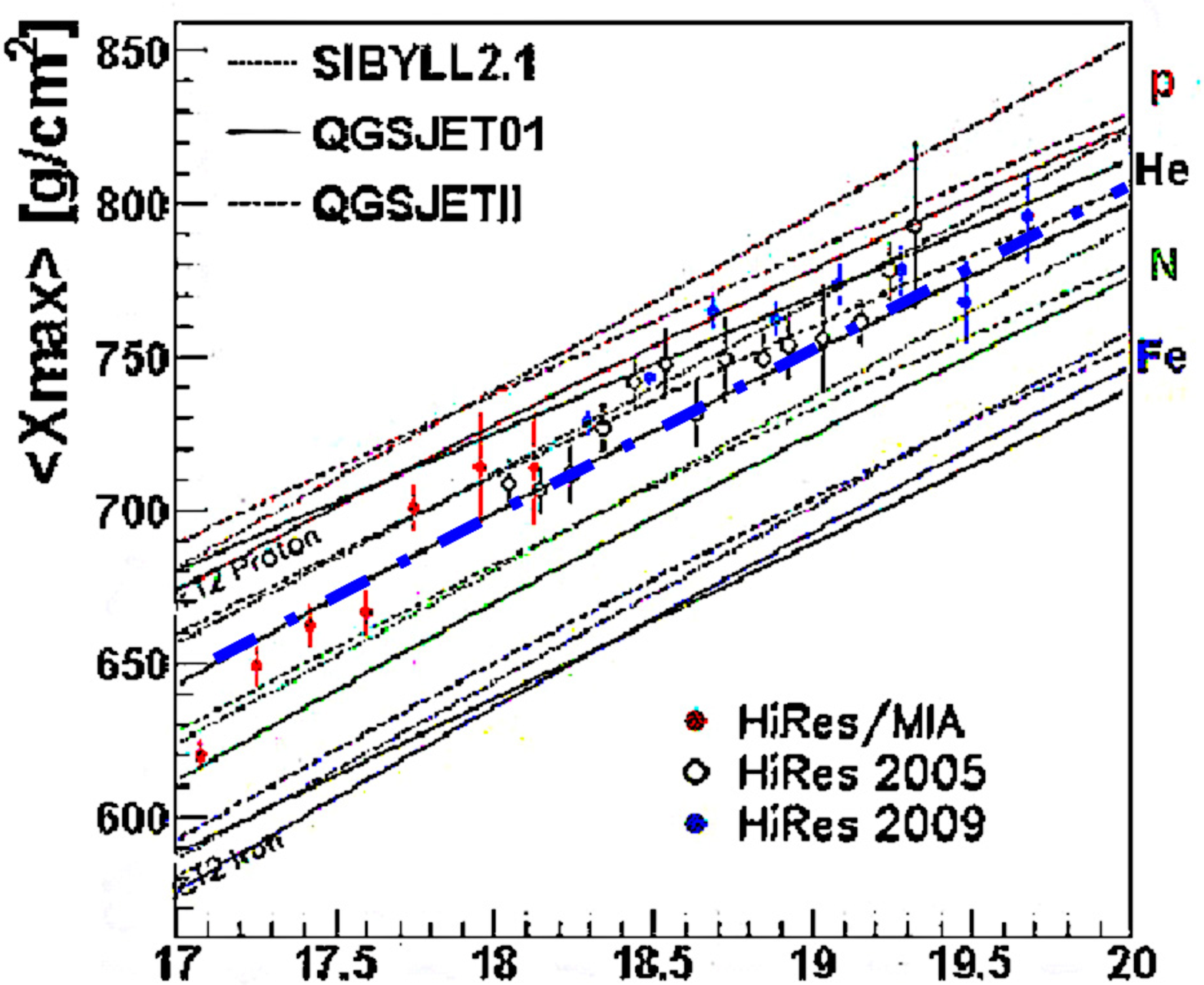}
   \includegraphics[width=3.2 in]{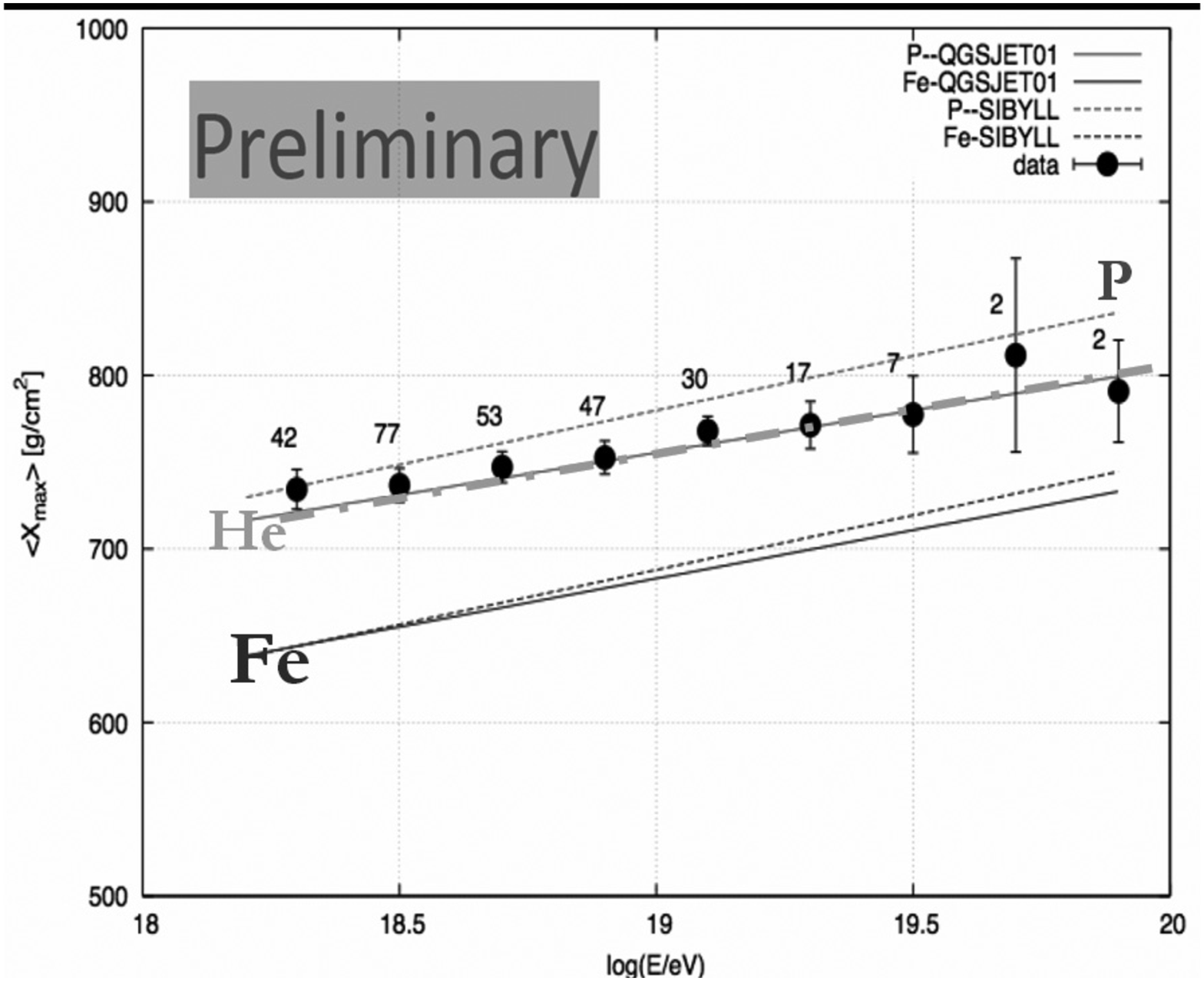}
  \caption{Left: one recent UHECR HIRES slant depth and composition derived by air shower feature; note the best fit of He on most of the highest UHECR events combining both Hires and AUGER results;
  right: one of  last UHECR Telescope Array Composition derived by air shower slant depth shown on 2011; note the best fit of He on most of the highest T.A. UHECR slant depth data}
  \label{fig3-4}
 \end{figure}

  Moreover the crowding of the two train multiplet tail (crosses center) inside a very narrow disk area focalized just along the rarest Cen A UHECR clustering  is remarkable (See Fig. \ref{fig2}).  If UHECR are  made by proton (as some AUGER author believe) they will not naturally explain such a wide tail structure because these proton events would not cluster more than a few degree, contrary to observed UHECR and associated multiplet (See Fig. \ref{fig2}). Also heavy nuclei whose smearing is much larger and whose eventual nucleon fragments ($A\rightarrow (A-1)$) should lead to collinear parasite tail at much lower energy,  greatly different  in mass and energy and bending angle from the observed AUGER ones. If UHECR are heaviest nuclei they should greatly spread and then only a nearest galactic smeared component may be somehow discovered (for the composition see Fig.\ref{fig3-4}) ( for gamma association map (See Fig. \ref{fig4-5},\ref{fig2})). Our dominant He-like UHECR do fit or compromise the AUGER and the HIRES composition traces (see Fig.\ref{fig3-4}). The He secondaries are splitting in two (or a fourth) energy fragments along Cen A tail (see Fig.\ref{fig2})  whose presence has  being foreseen and published many times  in last years\cite{Fargion2011}. Indeed the  dotted circle around Cen A containing the two (of three) multiplet centers  has a radius as small as $7.7^{o}$, (see Fig.\ref{fig2}),  it extend in an area that is as smaller as  $186$ square degrees, below or near $1\% $ of the observation AUGER sky (see Fig.\ref{fig2}). The probability that two among three sources of the multiplet fall inside such a  small Cen A area is offered by the binomial distribution. $$ P (3,2) = \frac{3!}{2!} \cdot (10^{-2})^{2} \cdot \frac{99}{100}\simeq 3 \cdot 10^{-4}$$ Moreover the same twin tail of the events are aligned almost exactly, within or less $\pm 0.1 $ rad parallel to UHECR train of events toward Cen A (see Fig.\ref{fig2}). Therefore the UHECR  multiplet clustering and alignment around Cen A  at twenty EeV has an a priori  probability as low as $ P (3,2) \simeq 3 \cdot 10^{-5}$ to occur following the foreseen signature\cite{Fargion2011}.

\section{Conclusions: UHECR fragments and UHE neutrinos}
 The history of Cosmic Rays and last UHECR discoveries (and disclaims) are exciting and surprising. The very unique correlation of UHECR with Cen A, the absence of Virgo, the hint of correlation with  Vela and a mild  connection with galactic plane or even Magellanic stream (see Fig.\ref{1r},\ref{fig2},\ref{fig2-3}) might be solved by a lightest nuclei, mainly He, as a  courier, leading to a very narrow (few Mpc) sky for UHECR.  A soon answer maybe already written into predicted \cite{Fargion2011} and now observed \cite{Auger11} multiplet clustering (as Deuterium or proton fragments) at half UHECR edge energy aligned  around or along main UHECR group seed discussed and shown above (see Fig.\ref{fig2}). Indeed He like UHECR  maybe bent by a characteristic as large as  $\delta_{rm-He}  \simeq 16^\circ$,(while expected  proton or D fragments at half fourth these energies, along tails spread  at $\delta_{rm-p}  \simeq 32^\circ$).  Future  AUGER map possibly about 2-3-4 tens EeV, UHECR fragment clustering maps along higher energy events (5-6 $10^{19}$ eV) may probe and reinforce our interpretation (already tested with present multiplet clustering at $ P (3,2) \simeq 3 \cdot 10^{-5}$ level). Additional clustering may occur along few galactic sources as Vela. More along galactic plane, as (inspired by Comptel gamma map correlation) with some clustering along Cassiopeia A, in future UHECR Telescope Array events  (see Fig.\ref{fig3-4}).  The discover of expected  Neutrino astronomy at Icecube or by horizontal  tau  neutrinos  airshower  at ARGO or  Auger,TA telescopes \cite{FarTau},\cite{Fargion1999}\cite{Fargion09b},\cite{Feng02},\cite{Auger08}, may also  shed  light on the UHECR nature, their origination  and mass composition,  opening our eyes to secret  UHECR sources.

\clearpage

\end{document}